# Magneto-Ionic Physical Reservoir Computing

Md Mahadi Rajib[1†], Dhritiman Bhattacharya[3†], Christopher J. Jensen[4], Gong Chen[3,5], Fahim F Chowdhury[1], Shouvik Sarkar[1], Kai Liu[3*], and Jayasimha Atulasimha[1,2,*]

[1]Department of Mechanical and Nuclear Engineering, Virginia Commonwealth University, Richmond, VA 23284, USA

[2]Department of Electrical and Computer Engineering, Virginia Commonwealth University, Richmond, VA 23284, USA

[3]Department of Physics, Georgetown University, Washington, DC 20057, USA

[4]NIST Center for Neutron Research, Gaithersburg, MD 20899, USA

[5]Department of Physics, Nanjing University, Nanjing, 210093, P. R. China

## Abstract

Recent progresses in magneto-ionics offer exciting potentials to leverage its non-linearity, short-term memory, and energy-efficiency to uniquely advance the field of physical reservoir computing. In this work, we experimentally demonstrate the classification of temporal data using a magneto-ionic (MI) heterostructure. The device was specifically engineered to induce non-linear ion migration dynamics, which in turn imparted non-linearity and short-term memory (STM) to the magnetization. These capabilities—key features for enabling reservoir computing—were investigated, and the role of the ion migration mechanism, along with its history-dependent influence on STM, was explained. These attributes were utilized to distinguish between sine and square waveforms within a randomly distributed set of pulses. Additionally, two important performance metrics—short-term memory and parity check capacity (PC)—were quantified, yielding promising values of 1.44 and 2, respectively, comparable to those of other state-of-the-art reservoirs. Our work paves the way for exploiting the relaxation dynamics of solid-state magneto-ionic platforms and developing energy-efficient magneto-ionic reservoir computing devices.

† Equal Contribution

*Email: kai.liu@georgetown.edu, jatulasimha@vcu.edu

## Introduction

Reservoir computing is a Recurrent Neural Network (RNN)-based framework [1, 2] that processes sequential and temporal data in a simple and efficient manner. In this scheme, signals that cannot be classified within the input space can be mapped to higher dimensions by leveraging the inherent nonlinearity of the reservoir layer, resulting in linearly separable outputs. Thus, the need to train multiple connections, as required in an RNN, is replaced by a reservoir layer. Outputs from the reservoir layer can simply be collected and multiplied by trained weights to perform classification or prediction tasks. These weights are not deep; they represent only a single neural network layer comprised solely of the reservoir outputs connected to the actual output for classification or prediction. Consequently, a simple linear regression model can be employed for training, making it simpler and more suitable for *in-situ* or online training compared to the more complex backpropagation. Furthermore, because this final layer consists of only a feedforward layer, it is not prone to issues like vanishing and exploding gradients, which affect RNNs and necessitate using Long Short-Term Memory (LSTM) and more recently transformers which take enormous energy and hardware to operate. All feedback connections in a reservoir are implemented with fixed connections, providing immense simplicity of the network and its training.

Central to the operation of a reservoir layer is the relaxation dynamics of a state parameter, which yields a parity check capacity reflecting nonlinearity and short-term memory, with past input memory retained but gradually fading due to damping effects. In a physical reservoir computer (PRC), the traditional reservoir layer is replaced by a physical system with such properties. For example, optical [3,4], memristive [5,6], mechanical [7, 8], and spintronic systems [9-14] have been demonstrated as the reservoir block of a physical reservoir computer. Among the various reservoir systems, spintronic reservoirs offer promising characteristics such as low-power consumption, scalability, and CMOS-compatibility [14, 15]. Various methods of controlling magnetization in spintronic devices, such as magnetic field [16], current [17-21], and electric field [22-31], have been demonstrated; among these, voltage-controlled methods have been shown to be the most energy-efficient [32, 33]. Most studies on voltage-controlled magnetism have focused on the control of interfacial magnetism at the metal/metal oxide interface, relying on the electric-field-induced change in the electronic structure of magnetic materials [22-27]. However, voltage control of ionic concentration at the interface of solid-state electrolyte and magnetic material has been shown to be highly energy-efficient [34-37], and the change in magnetic characteristics such as magnetic anisotropy energy can be 1-2 orders of magnitudes larger [29, 34]. However, despite the potential of leveraging such a highly energy-efficient method for controlling magnetization in a spintronic device, the demonstration of MI reservoir computing has yet to be achieved.

Here, we experimentally demonstrate that a MI heterostructure can be utilized for the implementation of physical reservoir computing. The nonlinear dynamics of ionic migration and corresponding change in magnetization in this platform are efficient for tasks such as temporal data classification. Particularly, we show classification of sine and square pulses from a pulse train consisting of randomly distributed sine and square pulses with 100% accuracy. Two important characteristic properties of a reservoir computer were also quantified, and promising values of 1.44 for STM and 2 for PC were found. For comparison, in a vortex-type ferromagnetic reservoir computing system, these values were found to be ~1.5 [38]. Another key advantage of MI devices is that their response time is typically in the range of milliseconds to minutes, which is similar to the time scale of many analog signals. It would be much more efficient to use MI reservoirs directly in such situations than to rely on complex electronics to convert signals to the ~1 GHz frequencies needed for implementing reservoir computing in spintronic devices, such as spin torque nano-oscillators (STNOs).

**Magneto-ionic reservoir system and STM property**

Figure 1 illustrates the framework of the MI physical reservoir computing system. In this setup, the reservoir layer is replaced by a MI heterostructure. A pulse train consisting of randomly distributed sine and square pulses was input into the MI device as voltage pulses. These voltage pulses induced ionic movement through the solid-state $GdO_x$ electrolyte, leading to changes in magnetization at the solid-state electrolyte/ferromagnet interface (refer to the Methods section for structural details). The resulting changes in magnetization were manifested in the hysteresis loops, collected using a magneto-optical Kerr effect (MOKE) microscope. The coercivity values obtained from the hysteresis loops were then trained with a linear regression model for waveform classification and for quantifying STM and PC capacity (see the Supplementary section for details on the STM and PC quantification methods).

In a solid-state MI device, a ferromagnetic layer is interfaced with a solid-state electrolyte through which ions, such as oxygen [34, 39, 40], hydrogen [36, 41], lithium [42], and nitrogen [43,44], can move upon the application of a voltage pulse. Depending on the polarity of the voltage pulse, the ions move either towards or away from the ferromagnetic layer, causing a change in magnetization [34]. Typically, these changes are non-volatile in nature, making them useful for implementing ultra-low power spintronic memory devices [34]. However, we designed our MI device with partially overlapped geometry, as shown in Fig. 1(b), where a volatile change offers the 'fading memory' property required to implement a reservoir computer. In this geometry, the oxygen ions can move laterally in the diffusion process, in addition to the vertical movement due to the electric field across the partially overlapping $GdO_x$ electrolyte region. This is discussed in detail later. The heterostructure's top electrode is grounded, and a voltage of ±8V is applied at the bottom electrode as shown in Fig. 2(a). When a positive voltage pulse is applied to a pristine heterostructure, oxygen ions moved towards the magnetic layer (Co layer), resulting in an increase in coercivity, as shown in Fig. 2(a). Conversely, when a negative voltage was applied, the oxygen ions moved away from the magnetic layer, causing a decrease in coercivity. As discussed earlier, these changes were volatile in nature, as illustrated in Figs. 2 (b-d).

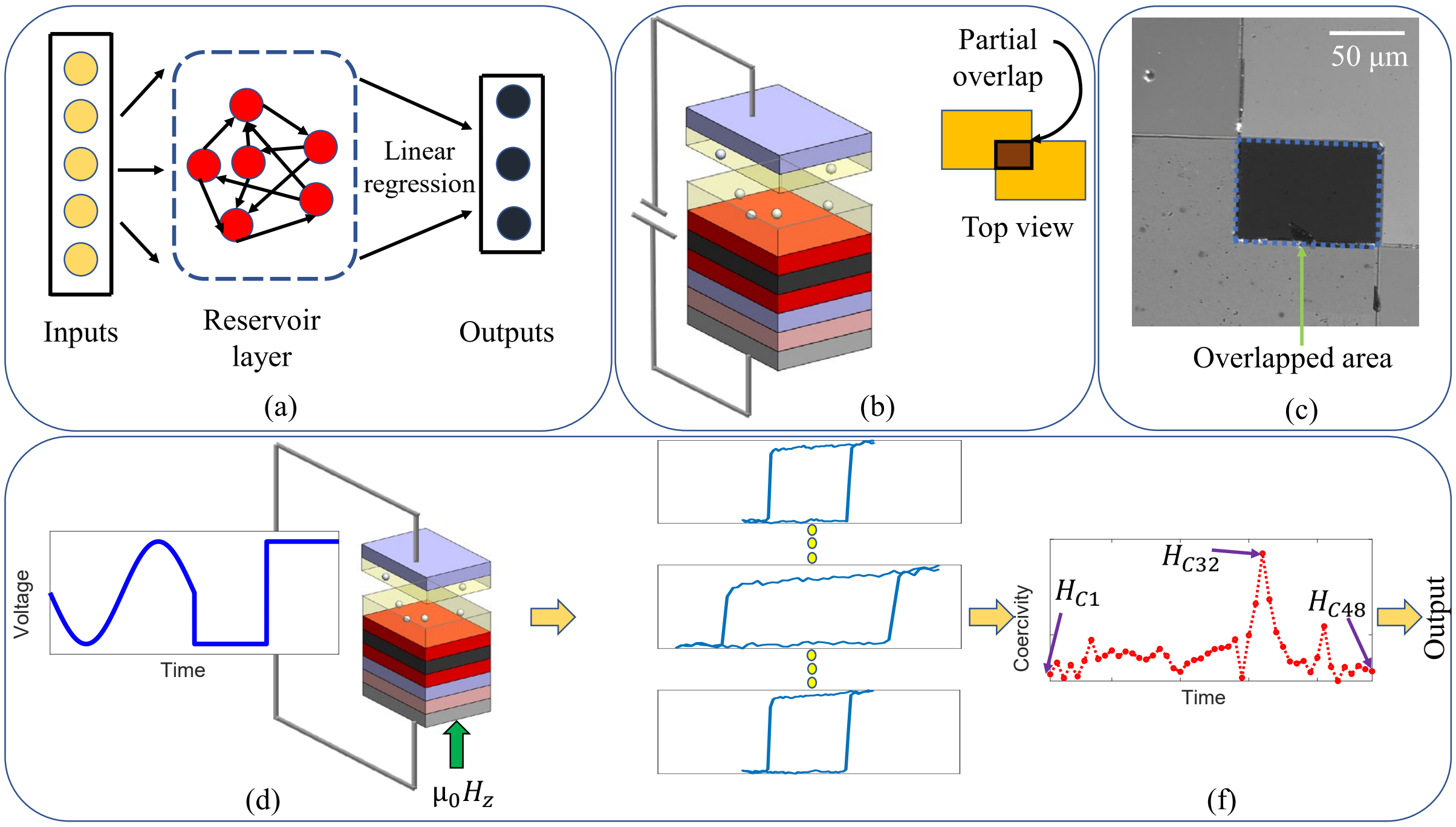

**Fig. 1: Magneto-ionic physical reservoir computing scheme.** (a) Schematic of a reservoir computing system based on a recurrent neural network: Inputs are mapped to a higher dimension via the reservoir layer, and outputs from the reservoir layer are classified using a simple linear regression model. (b) MI heterostructure as a reservoir block for physical reservoir computing: The inset depicts the partially overlapped geometry between the top and bottom electrodes. A MI PRC system is obtained by replacing the reservoir layer in (a) with the MI heterostructure. (c) MI heterostructure observed with a MOKE microscope: The area enclosed by the blue lines mark the region where changes in magnetization were manifested in hysteresis loops. (d) Schematic of the experimental setup: A sine waveform followed by a square waveform from a randomly distributed pulse train was applied to the MI heterostructure via voltage application. Magnetization changes due to the electric field across the $GdO_x$ were measured with MOKE by sweeping an out-of-plane (OOP) field along the perpendicular direction to the heterostructure, and hysteresis loops were recorded. (e) For each waveform, 48 hysteresis loops were recorded and labeled as $H_{c1}$, $H_{c2}$, ...$H_{c48}$. (f) Plot displaying 48 coercivity values for a sine pulse followed by a square pulse.

In the case shown in Fig. 2(b), as the voltage was initially decreased cumulatively from 0V to -20V, the coercivity also decreased monotonically (hysteresis loops for the initial state and -20 V shown in the inset). We note that the voltage was incremented by 2 V, with a 5-minute dwell time at each step, and it took approximately 1 minute to measure each hysteresis loop. Subsequently, the -20V voltage was removed and the coercivity was measured at different times. Interestingly, the coercivity began to increase from its minimum value upon removal of the -20V, eventually surpassing the initial coercivity of 7.17 mT, 80 minutes after the voltage was withdrawn (130 minutes after the voltage application began on a pristine heterostructure). The coercivity then continued to rise, reaching a maximum of 17 mT at 300 minutes after the withdrawal of the applied voltage (350 minutes after the voltage application began on a pristine sample), and stabilized at this value before gradually starting to decrease. While the coercivity decreased gradually at 0 V—by 2.4 mT over 400 minutes—the application of a positive voltage caused a more rapid reduction, with a 12.34 mT decrease observed as the voltage was cumulatively increased from 0 V to +20 V within 50 minutes. On a pristine heterostructure without prior gating, however, a purely positive pulse would increase coercivity, as shown in Fig. 2(a). These observations clearly demonstrate volatile relaxation dynamics with a pronounced history dependence in the magnetoelectric response, resulting in a nonlinear input response as well as short-term memory.

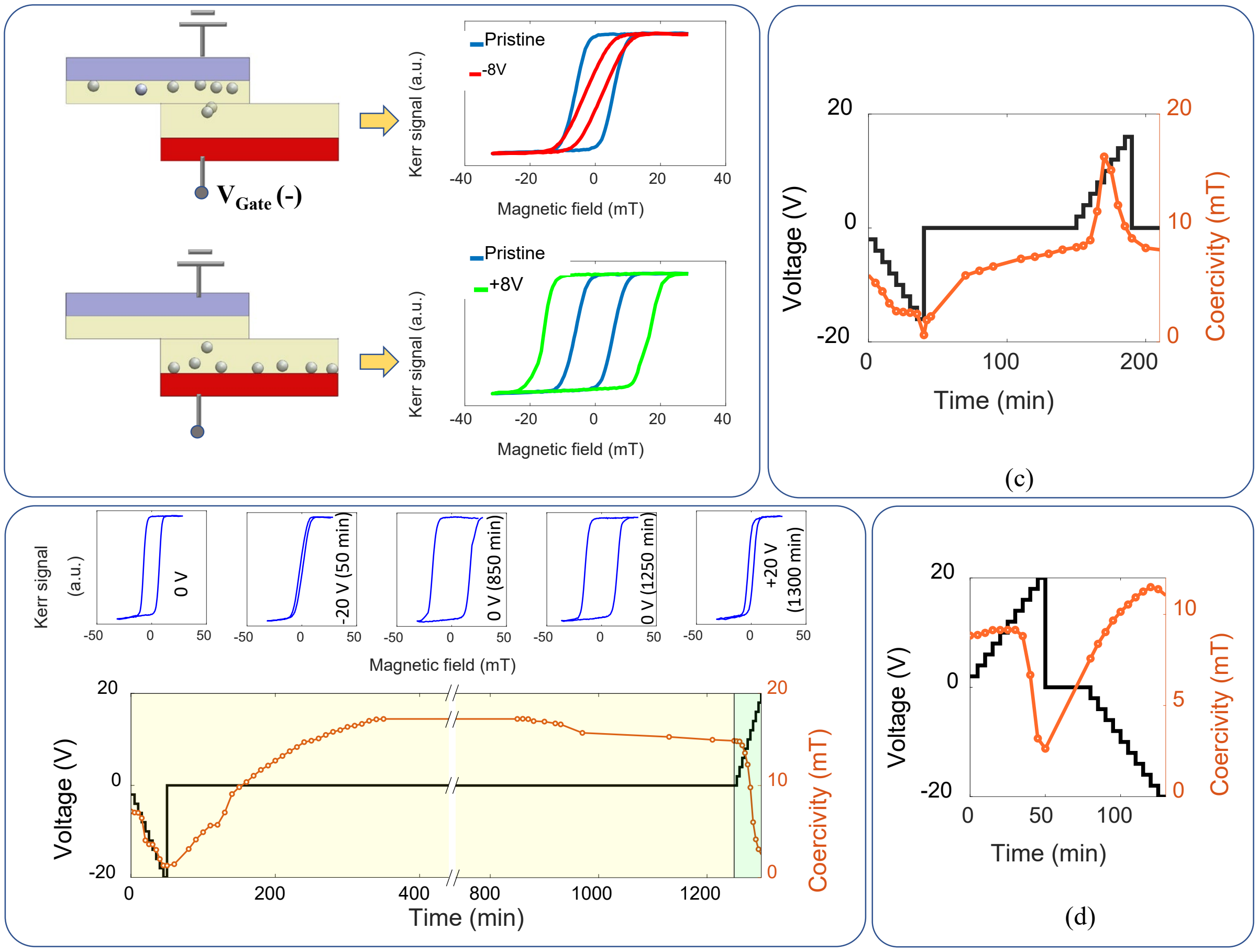

**Fig. 2: Relaxation dynamics and short-term memory.** (a) Change in magnetization in a Co/Ni thin film with positive and negative voltage pulse. A negative voltage pulse drives away oxygen ions from Co layer, reducing the coercivity, while a positive voltage pulse brings oxygen ions onto the Co surface, increasing the coercivity. (b, c, d) Volatile change in coercivity over time and demonstration of STM property. (Corresponding hysteresis loops are shown for 5 states in case (b)).

The scenario shown in Fig. 2(c) is similar to that in Fig. 2(b)—when negative voltage pulses were applied cumulatively to the heterostructure, the coercivity decreased, and upon withdrawal of the negative pulses, the coercivity began to increase. However, in this case, a positive voltage pulse was applied before the coercivity reached its peak, in contrast to the case shown in Fig. 2(b), where positive voltages were applied after the coercivity had already peaked. In Fig. 2(b), the coercivity increased from a minimum of 1.28 mT to a maximum of 17 mT over 300 minutes after the withdrawal of -20 V at a rate of 0.05 mT/min at 0 V, whereas in Fig. 2(c), the coercivity increased from a minimum of 0.61 mT to a maximum of 16.25 mT in 130 minutes. In Fig. 2(c), the coercivity increased at a rate of 0.06 mT/min at 0 V; however, the coercivity increased at a rate of 0.4 mT/min when positive voltages were increased cumulatively from 0 V to +10 V. This cumulative application of positive pulses in Fig. 2(c) accelerated the increase in coercivity, allowing it to reach the peak coercivity in a shorter time compared to the case in Fig. 2(b). After reaching the peak coercivity, further increases in the magnitude of the positive voltage up to +16 V led to a decrease in coercivity from 16.25 mT to 9.09 mT in 20 minutes, corresponding to a decrease rate of 0.36 mT/min.

Fig. 2(d) shows the case where a positive voltage pulses were applied first. Here, the coercivity increased with the cumulatively increasing positive voltage until it reached a peak of 9.15 mT at +10 V after 25 minutes, as shown in Fig. 2(d). Beyond this peak, further increases in the positive voltage to +20 V resulted in a decrease in coercivity to 2.64 mT. Upon withdrawal of the positive voltage pulses, the coercivity began to rise again. Subsequent application of cumulative negative pulses initially increased the coercivity to 11.5 mT after 120 minutes at -16 V, followed by a decrease to 11.0 mT after 130 minutes at -20 V.

In all three cases, it is evident that the effects of positive and negative voltages were influenced by the prior history of MI changes, unlike the independent effects observed on a pristine sample. In short, the data in Fig. 2 demonstrates that the MI heterostructure exhibits volatile relaxation dynamics, where the response to later pulses is governed by the history of earlier pulses, thereby imparting short-term memory properties to the system.

The dwell time was set on the minute scale due to the relatively slower voltage-driven ion migration at room temperature [34], as well as the time needed to measure a hysteresis loop corresponding to the applied voltage. The timescale for voltage-driven ion migration depends sensitively on both the amplitude of the voltage pulse and the temperature [45]. Therefore, applying higher voltages and temperatures can significantly shorten this timescale [34, 46]. Furthermore, modifying the oxide layer thickness allows for control over the oxygen ion migration distance, which in turn affects the timescale of the process. By adjusting these parameters, we can thus tune the response time of the MI device to align with the timescales required for specific reservoir computing applications.

**Physical origin of short-term memory in magneto-ionic device used**

Before utilizing these properties to implement reservoir functionalities, the underlying physical mechanisms behind the relaxation dynamics of the partially overlapping geometry were investigated. This was achieved by tracking ion migration through changes in optical contrast [47] using a MOKE microscope. The partially overlapping region of the device, outlined by a white dashed line in Fig. 3(a), is where the electric field is primarily concentrated. To examine ion migration, optical changes were monitored not only in the partially overlapping area but also in the adjacent top and bottom electrode regions. Note that, in our material system, a darker region indicates a shortage of oxygen ions whereas a brighter region indicates an abundance of oxygen ions [47].

To observe the migration of oxygen ions, voltage pulses were applied cumulatively. The voltage was reduced from 0 to -7V in -1V decrements per step, with a dwell time of 90 seconds at each step. Upon reaching the lowest voltage of -7V after 12 minutes, the bottom electrode area adjacent to the overlapped region darkened, as shown in Fig. 3(b). This is because when a negative voltage was applied to the bottom electrode, oxygen ions in the overlapping region were driven towards the top electrode due to the electric field across the heterostructure, while oxygen ions from the lateral bottom electrode area diffused into the overlapping region. As a result, the bottom electrode area near the overlapping region lost oxygen and appeared darker and the overlapping region became brighter. While the darker contrast in the bottom electrode area is readily recognizable in Fig. 3(b), the brighter overlapped area is not as apparent. To investigate further, we measured the intensity of the overlapped region associated with ion migration. The intensity of this region in Fig. 3(a) is 2008 (in arbitrary units), which increases to 2295 (or 14% brighter) after the application of -7 V. Note that such change in contrast in the bottom electrode area was not seen until -5V. This indicates that, due to the negative voltage pulses, oxygen ions were moving away from the magnetic layer, though significant lateral diffusion of ions from the adjacent bottom electrode area had not yet begun.

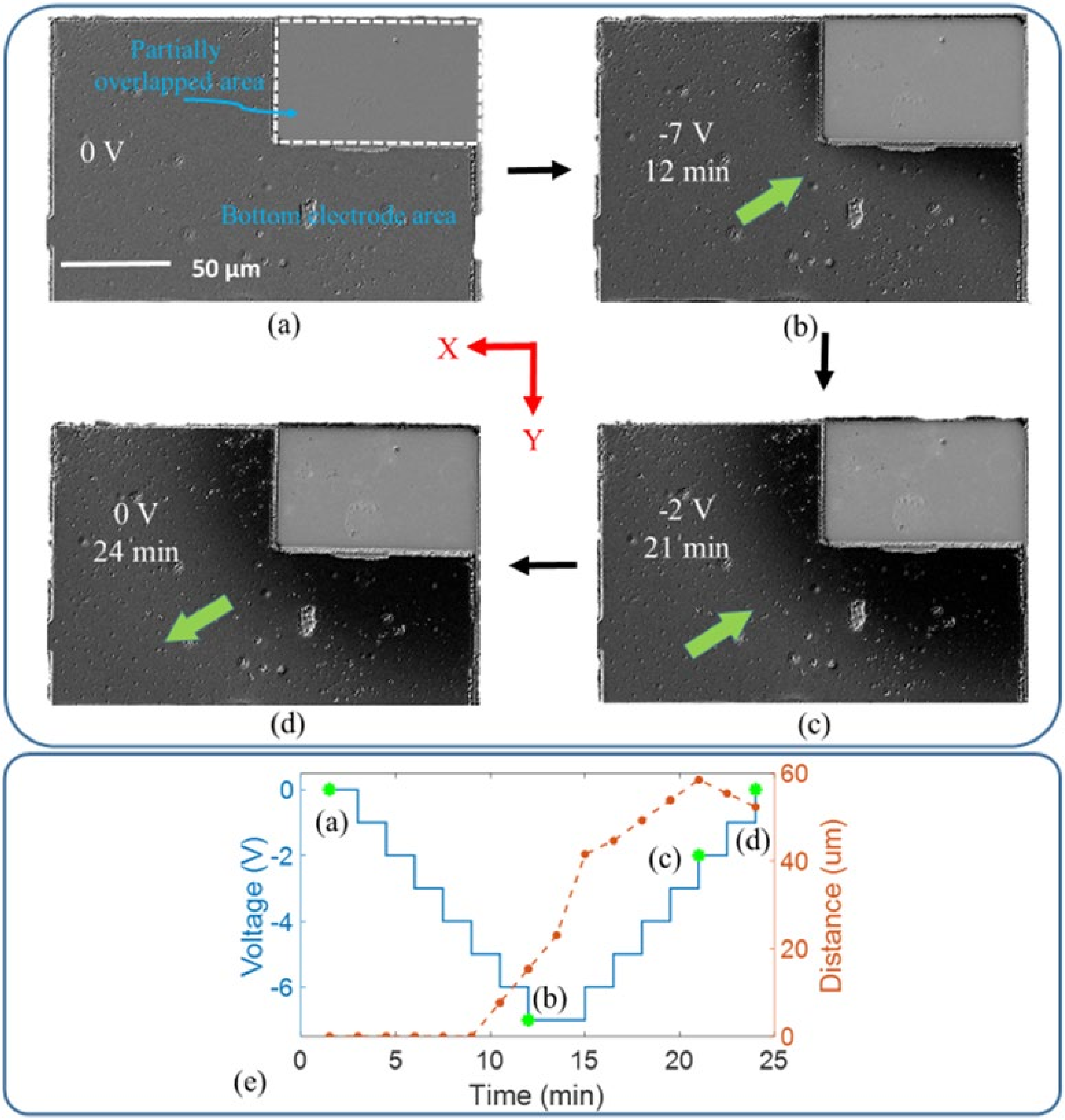

**Fig. 3: Physical explanation for short-term memory.** Migration of oxygen ions upon the application of voltage pulses: (a) Pristine state with the partially overlapping region marked by a white dashed line. (b) Oxygen ions move inside the overlapping area, causing that area to become brighter due to the vertical movement of oxygen ions towards the top electrode and the diffusion of oxygen ions from the lateral bottom electrode area. (c) More oxygen ions enter the overlapping area, making it much brighter, while other areas of the lateral bottom electrode become darker. (d) The concentration of oxygen ions after the withdrawal of the voltage pulses differs from that before the application of voltage pulses in (a). (e) Voltage vs distance traveled by the oxygen ions along the X-axis.

Subsequently, as the voltage was increased cumulatively, even at a lower negative voltage (-2V), additional oxygen ions continued to diffuse toward the overlapping area, causing further darkening in the lateral bottom region while the overlapping region brightened (Fig. 3c) with an intensity of 2301. After the negative voltages were withdrawn, the oxygen ion concentrations in both the overlapping and adjacent bottom electrode areas differed (Fig. 3d) compared to their initial levels before the voltage pulse (Fig. 3a), indicating that ion migration was not directly proportional to the applied voltage. The difference in optical contrast between the lateral bottom electrode areas in Fig. 3(a) and 3(d) is readily recognizable, while the difference in optical contrast in the overlapped region can be inferred from the intensities, which are 2008 and 2251 for Fig. 3(a) and Fig. 3(d), respectively. Although oxygen ion migration occurs in three dimensions, measurements were taken along one axis to track the migration distance. As seen in Fig. 3(e), the migration distance varied non-linearly with the applied voltage. This behavior contributed to the short-term memory property of the reservoir.

**Reservoir task**

Next, we demonstrate the reservoir task. During magnetic field and voltage cycling, the MOKE intensity was influenced by both ionic migration and magnetization orientation. Therefore, it was not used directly for reservoir task demonstration. Instead, coercivity was used as a more direct measure of the magnetic properties of the system. Specifically, the coercivity changes in the partially overlapping area of the device in response to randomly distributed sine and square waveforms were measured and analyzed. Figure 4(a) illustrates an example of coercivity variation in response to sine and square waveforms. We observe a gradual increase in coercivity as the pulses are successively applied, potentially caused by the irreversible increase in ion concentration that occurs during the positive cycles of the pulses. The pulse width of the input voltage pulses was set to 36 minutes, and 24 hysteresis loops were measured at 90-second intervals for each waveform. Fig. 4(b) presents a zoomed-in view of the green region from Fig. 4(a), showing the coercivity changes between the 27th and 30th pulses. It was observed that for square pulses, the coercivity plot exhibited sharper peaks compared to sine pulses.

From these results, the memory capacity of the MI system was examined, by determining whether the current magnetization response was influenced by previous pulses, i.e., whether the magnetization response differed for various combinations of past and present pulses. Figure 4 illustrated the coercivity for the "present waveforms," which had a known past input, showing that the magnetization responses were identical for the same combination and different for different combinations of waveforms. Four possible past-present pulse combinations of two types of pulses (sine and square) were considered: square-square, sine-sine, sine-square, and square-sine, where the first pulse in each combination represents the past pulse, while the latter represents the present pulse. The response to the "present" pulse from each combination was plotted to assess how the past pulse altered the response of the present pulse. In Figures 4(c), 4(e), and 4(d), 4(f), the present pulses were square and sine, respectively. These figures display the polyfitted normalized coercivity vs timesteps. Each of the red lines in Fig. 4(c)-(f) was plotted using 24 normalized hysteresis loops (time steps), with each red line representing the output for a single pulse (either sine or square). The multiple curves in each panel correspond to individual pulses from the randomly distributed pulse train shown in Fig. 4(a). These curves are sorted based on the input 'past-present' pulse combinations, and the curves are nearly similar in each of the corresponding panels in Fig. 4(c)-4(f).

When the square waveform was applied as the present pulse, the coercivity plot exhibited a wider dome shape [see Fig. 4(c) and 4(e)], while for the sine pulse, the domes were narrower [see Fig. 4(d) and 4(f)]. However, the magnetization response varied in shape depending on the past input, although the dominant feature resulting from the present input was retained. In Fig. 4(c) and 4(e), the normalized coercivity for a square pulse as the present input is displayed, with a square pulse as the past input in case (c) and a sine pulse in case (e). Indeed, the response for the present square pulse varied in shape depending on the past pulse. When the past pulse was square, the two ends of the dome-shaped normalized coercivity became flat (indicated by a flat dotted blue line), whereas, with a past sine pulse, the ends of the dome took on a slanted shape (indicated by a slanted blue dotted line). Despite these differences, both cases maintained the wider dome shape, which was the dominant feature caused by the present square pulse, while the orientation of the dome varied depending on the past input.

Similarly, for 4(d) and 4(f), the present pulse was sine, which resulted in a narrow dome shape as the dominant feature in both cases, with the peak of the dome occurring at the 10th time step of the pulse for past sine pulses. In contrast, the peak was observed at the 8th time step for past square pulses. These observed differences in magnetization response for varying combinations, along with the similarities for the same combination without overlap, suggested that the MI heterostructure exhibited memory and nonlinearity.

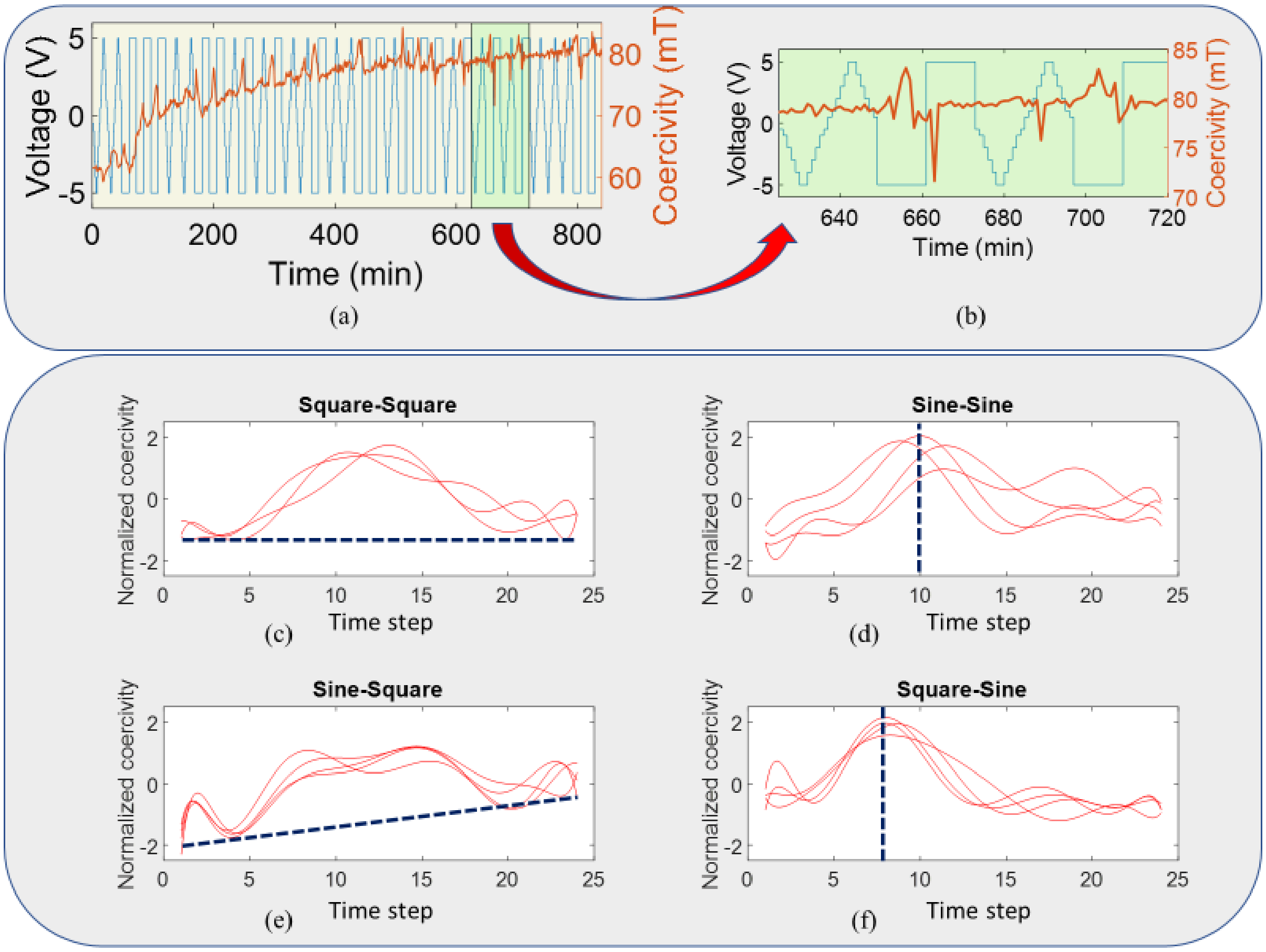

**Fig. 4: Demonstration of reservoir task.** (a) Coercivity changes induced by randomly distributed sine and square voltage pulses. (b) Coercivity changes for pulses 27th to 30th are displayed in the inset (marked by green color in (a)). A distinct pattern emerges in the coercivity changes observed for sine and square pulses, with the square pulses demonstrating sharper peaks in comparison to the sine pulses. The normalized coercivity response for a "present" pulse from a pulse combination of (c) square-square, (d) sine-sine, (e) sine-square, and (f) square-sine.

These results were input into a simple linear regression model. When training was conducted on 31 input pulses, and testing was performed on 4 pulses, the MI system recognized sine and square pulses with 100% accuracy. This indicated that the MI system was capable of distinguishing between sine and square waveforms within a randomly distributed set of such pulses. Additionally, the STM and PC capacities were also quantified. The STM and PC capacities represent the number of data points stored for linear and nonlinear combinations of input data, respectively [38]. By considering a delay of D=1, STM and PC capacities of the MI device for 24 virtual nodes were found to be 1.44 and 2, respectively, which were comparable to those of other state-of-the-art reservoirs [38, 48-50]. We note that in a single node reservoir, time multiplexing was utilized to mimic a reservoir [9, 51], as was the case with the single MI device in our experiments.

**Discussion and Conclusion**

In summary, we have demonstrated that the volatile relaxation dynamics of ionic migration in a MI device with a partially overlapping geometry endow it with short-term memory properties, making it suitable for reservoir computing. This MI reservoir is capable of performing temporal data classification tasks with a small number of training datasets, achieving 100% accuracy. Notably, our system exhibits two key characteristic properties of a reservoir computer: STM and PC capacity, with promising values of 1.44 for STM and 2 for PC, respectively. The demonstrated reservoir, with a comparatively larger timescale, can be readily useful in predicting time-series such as household energy load [12] and weather forecasting [52]. However, this timescale can be further reduced by selecting a different MI system or mechanism [53], depending on the type of task. In our experiments, the use of MOKE to measure hysteresis loops and capture changes in magnetization in response to applied voltage pulses limited the timescale we could probe. Utilizing a Hall bar or a magnetic tunnel junction to measure the change in magnetoresistance due to alterations in magnetization would significantly improve this timescale. Furthermore, by applying higher voltages and increasing temperature, a super-exponential transition in voltage-driven ion migration occurs [45], resulting in a reduction of the timescale from several minutes to several microseconds [34]. Although our proposed MI reservoir, with its timescale, is advantageous for certain specific types of tasks, the reduction in timescale with voltage and temperature, combined with an appropriate choice of MI platform, could create opportunities for implementing task-adaptive MI reservoirs. Hence, our work lays the groundwork for energy-efficient MI reservoir computing.

**Method**

For fabricating the devices, a heterostructure consisting of Ta (3)/Cu (10)/Pd (3)/Co (0.7)/Ni (0.3)/Co (0.7)/$GdO_x$ (10) (all numbers in nm), was grown on Si/$SiO_2$ and patterned into 300 µm squares using standard photolithography and lift-off technique. Next, $GdO_x$ (15)/Pd (5) layers were grown using the same 300 µm square mask. This square pattern was shifted with respect to the first pattern to create the partially overlapped region of 50 µm, as illustrated in Fig. 1(b). All layers were sputtered using DC magnetron sputtering with a base pressure higher than $3\times10^{-8}$ Torr. The metallic layers were sputtered with Ar working pressure of 2.5 mTorr. The $GdO_x$ layer was reactively sputtered with a mixture of Ar and $O_2$ (77:2.9) with a working pressure of 5 mTorr.

An evicomagnetics MOKE microscope was used to observe the change in magnetization and measure the hysteresis loops. The polar mode was utilized to acquire the coercivity data. Hysteresis loops were obtained by applying an OOP field with a step size of 0.2 s, and each loop took 40 to 90 seconds to measure, depending on the range and the chosen step size of the applied magnetic field. Voltage pulses were applied to the heterostructure in situ using a Keithley 2636B source meter, while measuring the hysteresis loops with the MOKE microscope.

## Acknowledgements

This work has been supported in part by the NSF (CCF-1909030, DMR- 2005108 and ECCS-2151809) and Commonwealth Cyber Initiative.

## Data availability statement

The data that supports the findings of this study are available upon reasonable request from the authors.

## Disclaimer

Certain equipment, instruments, software, or materials are identified in this paper in order to specify the experimental procedure adequately. Such identification is not intended to imply recommendation or endorsement of any product or service by NIST, nor is it intended to imply that the materials or equipment identified are necessarily the best available for the purpose.

Supplementary Information

# Magneto-Ionic Physical Reservoir Computing

Md Mahadi Rajib[1†], Dhritiman Bhattacharya[3†], Christopher J. Jensen[4], Gong Chen[3,5], Fahim F Chowdhury[1],

Shouvik Sarkar[1], Kai Liu[3*], and Jayasimha Atulasimha[1,2,*]

[1]Department of Mechanical and Nuclear Engineering, Virginia Commonwealth University, Richmond, VA 23284, USA

[2]Department of Electrical and Computer Engineering, Virginia Commonwealth University, Richmond, VA 23284, USA

[3]Department of Physics, Georgetown University, Washington, DC 20057, USA

[4]NIST Center for Neutron Research, Gaithersburg, MD 20899, USA

[5]Department of Physics, Nanjing University, Nanjing, 210093, P. R. China

## Classification and STM/PC quantification method

We use a pulse train made up of 35 randomly distributed sine and square pulses, each with a time period of 36 minutes, to demonstrate simple temporal pattern recognition tasks. The initial 31 pulses serve as training data, while the last 4 are reserved for testing the reservoir computer. The randomly distributed sine and square pulses are applied to the reservoir block, where they are labeled as 1 and 0, respectively, for the classification task:

$$p(i)=\begin{cases} 1 \text{ for sine pulse} \\ 0 \text{ for square pulse}, \end{cases} \tag{1}$$

$$p(i) = \{p_1, p_2, \ldots\ldots\ldots, p_T, p_{T+1}, \ldots., p_F\} = \{1,1,0, \ldots\ldots..,1,1\} = \{X_T, X_P\}, \tag{2}$$

here, $i$ denotes the index of the input pulse, where $i \in \{1,2,3, \ldots\ldots.T, \ldots., F\}$. $X_T$ and $X_P$ represent the training and testing datasets, respectively:

$$X_T = \{p_1, p_2, \ldots\ldots\ldots, p_T\},$$

$$\text{And } X_P = \{p_{T+1}, p_{T+2}, \ldots\ldots\ldots, p_F\},$$

Here, "F" represents the total number of pulses, and "T" indicates the number of pulses used for training.

The output (coercivity) of the reservoir block is expressed as an N×F matrix:

$$H_{mn} = \begin{bmatrix} h_{11}\ h_{12}\ \dots\ h_{1F} \\ h_{21}\ h_{22}\ \dots\ h_{2F} \\ \dots.\ \dots.\ \dots\ \dots \\ h_{N1}\ h_{N2}\ \dots\ h_{NF} \end{bmatrix}; \quad (3)$$

$m \in \{1,2,3, \dots\dots\dots.., N\}$ and $n \in \{1,2,3, \dots\dots\dots.., F\}$,

N represents the virtual node [1], defined as $N=\frac{P}{\tau}$, where $P$ is time period of the pulses, and $\tau$ is the time interval at which the output (coercivity) is measured. The concept of virtual nodes was introduced by Y. Paquot et al. [1]. By connecting virtual nodes sequentially in time and feeding them back into the nonlinear node, a specific type of recurrent neural network (RNN) architecture is created [2, 3]. In this context, virtual nodes correspond to the number of coercivity values within each time period.

The weight vector is determined by:

$$W=X_T \times pinv(H_{mn}), \quad (4)$$

m $\in$ {1,2,3, ..........., N} and n $\in$ {1,2,3, ..........., T}as we consider the training data only.

Here *pinv* is the Moore-Penrose Pseudoinverse of matrix $H_{mn}$

The test data can be reconstructed by using the weight learnt during training:

$$X_{RO} = W \times H_{mn}, \quad (5)$$

m $\in$ {1,2,3, ..........., N} and n $\in$ {T+1, T+2, ..........., F} as we consider the testing data only.

The pattern recognition task evaluates whether the reconstructed outputs ($Y_{RO}$) can accurately predict the test data ($X_P$).

As mentioned earlier, in addition to checking accuracy, the reservoir's performance can also be measured through its STM and PC capacities. STM capacity reflects the reservoir's ability to reconstruct past inputs based on its present outputs. The input training and testing data with delay a D ($X^{STM}_{Train,n-D}$, $X^{STM}_{Test,n-D}$) for calculating STM capacity are determined as follows:

$$X_{STM} = p(i-D) = \{X^{STM}_{Train,n-D}, X^{STM}_{Test,n-D}\}, \quad (6)$$

Conversely, PC capacity evaluates the nonlinear transformation ability of a reservoir block. The input training ($X^{PC}_{Train,n-D}$) and testing ($X^{PC}_{Test,n-D}$) data for determining PC capacity are derived using a modulo operation as follows:

$X_{PC}$=[p(i-D) + x(i-D+1) +.... + p(i)] mod (2) = $\{X^{PC}_{Train,n-D}, X^{PC}_{Test,n-D}\}$, (7)

To evaluate STM and PC capacities, the capability of the reconstructed output to predict the test data with delay D is estimated from the following correlation coefficient [4]:

$$Cor(D) = \frac{\sum_{k=1}^{Z}(X_{Test,n-D} - < X_{Test,n-D} >)(X_{RO,n-D} - < X_{RO,n-D} >)}{\sqrt{\sum_{k=1}^{Z}(X_{Test,n-D} - < X_{Test,n-D} >)^2 \sum_{k=1}^{Z}(X_{RO,n-D} - < X_{RO,n-D} >)^2}}, \quad (8)$$

Here, $X_{Test,n-D}$ and $X_{RO,n-D}$ denote the test data and reconstructed output at a delay D, respectively, and <···> represents the mean value of "Z" number of data.

STM or PC capacity is calculated using the following equation [4]:

$$C_{STM/PC} = \sum_{D=1}^{D'} [Cor(D)]^2, \quad (9)$$

Where $D'$ is the maximum delay.